\documentclass[%
 reprint,
 amsmath,amssymb,
 aps,
]{revtex4-2}

\usepackage{graphicx}
\usepackage{dcolumn}
\usepackage{bm}

\usepackage{braket}
\usepackage{subfigure}

\begin{document}

\title{Non-Abelian statistics with mixed-boundary punctures on the toric code}

\author{Asmae Benhemou$^1$, Jiannis K. Pachos$^2$ and Dan E. Browne$^1$}
\affiliation{$^1$Department of Physics and Astronomy, University College London, London WC1E 6BT, United Kingdom}
\affiliation{%
 $^2$School of Physics and Astronomy, University of Leeds, Woodhouse Lane, Leeds LS2 9JT, UK}%

\date{\today}

\begin{abstract}
The toric code is a simple and exactly solvable example of topological order realising Abelian anyons. However, it was shown to support non-local lattice defects, namely twists, which exhibit non-Abelian anyonic behaviour \cite{Bombin_2010}. Motivated by this result, we investigated the potential of having non-Abelian statistics from puncture defects on the toric code. We demonstrate that an encoding with mixed-boundary punctures reproduces Ising fusion, and a logical Pauli-$X$ upon their braiding. Our construction paves the way for local lattice defects to exhibit non-Abelian properties that can be employed for quantum information tasks.  
\end{abstract}

\maketitle


\section{\label{sec:level1}Introduction}

Anyons are excitations in two-dimensional systems, that are neither bosons nor fermions \cite{PhysRevLett.49.957}. Abelian anyons collect an arbitrary complex phase factor upon exchange. The exchange of two non-Abelian anyons is described by a matrix representation of the braid group \cite{Freedman} acting on the Hilbert space describing the composite anyonic system. The latter type is of particular interest since its anyons can be used to process information by braiding them in a topological quantum computing scheme \cite{pachos_2012,preskill}. Anyons emerge in phases of matter that have topological order such as Fractional Quantum Hall (FQH) states, the Kitaev honeycomb lattice model (KHLM), quantum double models \cite{Kitaev_2003, pachos_2012} etc. The Ising model famously characterises the behaviour of quasi-particles arising from physical systems supporting Majorana Zero Modes (MZM) \cite{Kitaev_2001,Fu_2008}.  

Lattice models consisting of a qubit ensemble arranged on a two-dimensional surface are a practical tool to study such topological systems. These models, such as stabiliser codes \cite{phdthesis, Fowler_2012}, allow for computational schemes that encode quantum information in non-local degrees of freedom. The canonical example is the toric code, introduced by Kitaev in ref.~\cite{Kitaev_2003}. It encodes logical qubits in the degenerate ground states of a square spin lattice defined on a torus \cite{Kitaev_2006}. The toric code was shown to emerge in the Abelian phase of the KHLM \cite{Kitaev_2006, Kells_2009}. 

The toric code admits local, point-like defects and non-local, line-like defects. Punctures are local defects corresponding to holes on the lattice. They were introduced as candidates for quantum memory and computation through their braiding \cite{Dennis_2002,  delfosse2016generalized, Raussendorf_2007}, while twists are the endpoints of non-local domain walls that enforce a symmetry on the toric code anyons. The latter defects have been described with topological quantum field theories (TQFTs) \cite{Barkeshli_2013, Teo_2016}. They are also computationally interesting since they were shown to behave like Majorana zero modes under fusion and exchange \cite{Bombin_2010, Zheng_2015, You_2012}. A novel hybrid of these two defect types was even introduced in ref.~\cite{Krishna_2020}, also capable of encoding logical qubits.

In this article we investigate the topological properties of yet another defect on the toric code, namely punctures with mixed-boundaries. Following the adiabatic equivalence between vortices and twists demonstrated in the non-Abelian phase of Kitaev’s honeycomb lattice model \cite{horner}, we studied all possible deformations of twists that could be adiabatically deformed to punctures. None of them worked in providing point-like defects that could support the desired statistics. Thus, we resorted to the mixed-boundary punctures as the optimal tool for an encoding of Majoranas that takes advantage of the Abelian statistics and non-local encoding of gates. Moreover, our choice is congruent with the insight in ref.~\cite{Wootton_2008} that the quantum dimensions of the Ising model and of the toric code are equal. This gives the basis for employing toric code anyonic statistics in order to realise more complex Ising anyon properties.
In particular, we demonstrate non-Abelian fusion and braiding properties reminiscent of Majorana exchange. To achieve it we employ local lattice defects of the toric code with mixed boundary conditions in conjunction with a non-local logical encoding between them. Our approach enriches the type of defects that can reproduce the behaviour of Majorana anyons, thus helping to close the gap between their exotic statistics and their physical realisation or possible simulation with a quantum computer.

This work is organised as follows; In section \ref{sec:background} we review the planar code defects and briefly outline non-Abelian Ising statistics and their relation to twist defects. In section \ref{sec:naStats} we introduce defects which generalise punctures to ones with mixed-boundary conditions in order to encode non-Abelian fusion rules. We demonstrate their Ising-like fusion and braiding statistics after defining a
logical encoding based on a superposition of their
population states. We conclude and discuss our results in section \ref{sec:conclusion}.

\section{\label{sec:background} Background}

\subsection{Anyon models}

Anyon models are algebraic structures that characterise topological order in many body systems. They comprise excitations $a,b,... \in \mathcal{C}$, where $\mathcal{C}$ is a finite set of quasi-particles, distinguished by their charges. Two anyons $a$ and $b$ fuse following $a \times b = \sum_{c}N_{ab}^c c$, where $N_{ab}^c$ is the multiplicity of outcome $c \in \mathcal{C}$. Moreover, the braiding rules of an anyon model are specified by the phases or operators obtained under their exchange. The toric code is a quantum double of $\mathbb{Z}_2$ with Abelian anyons that comprises a vacuum charge \textbf{1}, excitations $e$, $m$, and their composition $\epsilon$. All of the above fuse to vacuum when composed with an anyon of the same type, and also obey $e \times m = \epsilon$. The self and mutual statistics of toric code anyons are described by the $R$-matrices, i.e. the evolution operators describing the exchange of anyons
\begin{equation} 
\begin{aligned}
    R_{ee} &= R_{mm} = 1 \\
    R_{\epsilon\epsilon} &= -1 \\
    R_{em}R_{me} &= R_{e\epsilon}R_{\epsilon e} = -1.
    \label{eq:toricstats}
    \end{aligned}
\end{equation}
The braiding relations in Eq.~\ref{eq:toricstats} tell us that $e$ and $m$ are both bosons, while $\epsilon$ is a fermion. Additionally, $e$ and $m$ are mutual semions, meaning that braiding an $e$ around an $m$ returns a phase of $-1$, and similarly for $e$ and $\epsilon$ (and $m$ and $\epsilon$). In a surface code defined on a lattice of qubits, $e$ and $m$ anyons emerge at the ends of strings of respectively $Z$ and $X$ operations.

In the following we will refer to defects  on the toric code which have similar characteristics and behaviour as Ising anyons. These belong to the non-Abelian Ising model which is characterised by anyonic charges \textbf{1}, $\sigma$ and $\psi$ such that
\begin{equation}
    \psi \times \psi = \textbf{1}, \hspace{2mm} \psi \times \sigma = \sigma, \hspace{2mm} \sigma \times \sigma = \textbf{1} + \psi
    \label{eq:isingmodel}
\end{equation}
where $\psi$ is a fermion, and two Ising anyons $\sigma$ can fuse to the vacuum charge \textbf{1} or one $\psi$. Hence, there is a two-dimensional Hilbert space associated with a pair of $\sigma$ anyons, with basis states characterised by the fusion channels, i.e
\begin{equation}
\ket{\sigma\sigma \xrightarrow{} \textbf{1}} \hspace{2mm}\text{and}\hspace{2mm}
 \ket{\sigma\sigma \xrightarrow{} \psi}.
\end{equation}
In order to access superpositions of these two states a qubit can be encoded in the global state of a composite system of four $\sigma$ anyons, under the constraint that the total fermion parity is conserved. The basis in this space can be spanned by 
\begin{equation}
\ket{(\sigma\sigma)(\sigma\sigma) \xrightarrow{} \textbf{1};\textbf{1}} \hspace{1mm}\text{and}\hspace{1mm}
\ket{(\sigma\sigma)(\sigma\sigma) \xrightarrow{}\psi;\psi}.
\label{fig:basissector}
\end{equation}
Let us assume we have four Ising anyons enumerated 1, 2, 3 and 4. Modifying the fusion order from $(12)(34)$ to $(13)(24)$ corresponds to a basis change in this Hilbert space given by a fusion matrix 
\begin{equation}
F_{\text{Ising}} = \frac{1}{\sqrt{2}}
\begin{pmatrix}
    1 & 1 \\
    1 & -1 
\end{pmatrix}
\label{eq:fmatrix}
\end{equation}
while all other F-matrices in the model are phases only \cite{pachos_2012}. Moreover, the non-trivial braiding relations are given by 
\begin{equation} 
\begin{aligned} 
R_{\psi\psi} &= -1, \qquad
R_{\psi\sigma}R_{\sigma\psi} = -1 \\
R_{\sigma\sigma} &= e^{-i\frac{\pi}{8}}
\begin{pmatrix}
    1 & 0\\
    0 & i
\end{pmatrix}. 
\end{aligned}
\end{equation}
The braiding evolution for Ising anyons is described using the above $F$ and $R$ matrices such that
\begin{equation}
    B = FR^2F^{-1} = e^{-i\frac{\pi}{4}} \begin{pmatrix}
        0 & 1 \\
        1 & 0
    \end{pmatrix}
    \label{eq:anyonX}
\end{equation}
which is a non-trivial unitary logical operation, specifically the Pauli-$X$ gate (up to a global phase factor). This feature gives rise to the Clifford group by braiding anyons from two $\sigma$-pairs encoding a qubit. This model is reproduced by Ising anyons in Fractional quantum Hall states, as well as Majorana zero-modes in topological superconductors \cite{ivanov, Litinski_2018}.

\subsection{Toric code defects}

\begin{figure}
    \centering
    \subfigure(a){\includegraphics[width=0.20\textwidth]{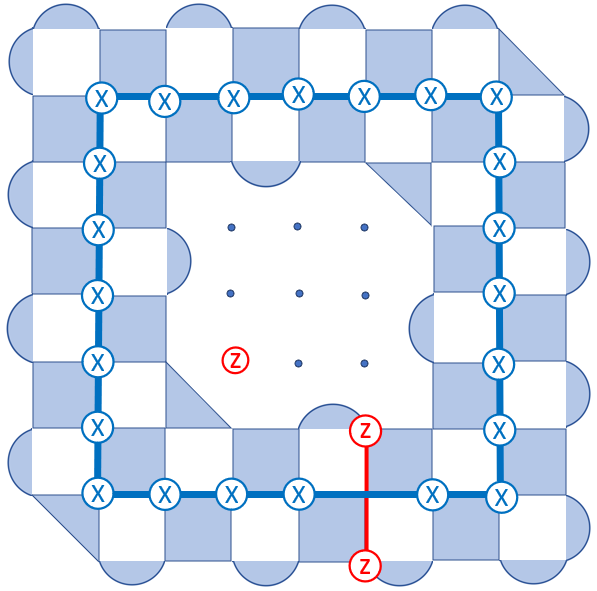}} 
    \subfigure(b){\includegraphics[width=0.19\textwidth]{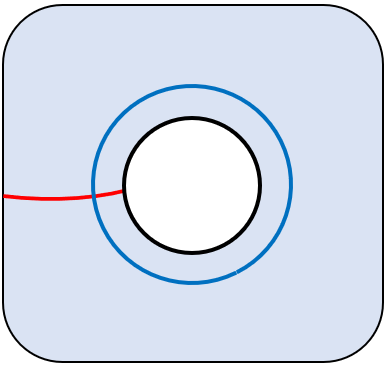}} 
     \subfigure(c){\includegraphics[width=0.20\textwidth]{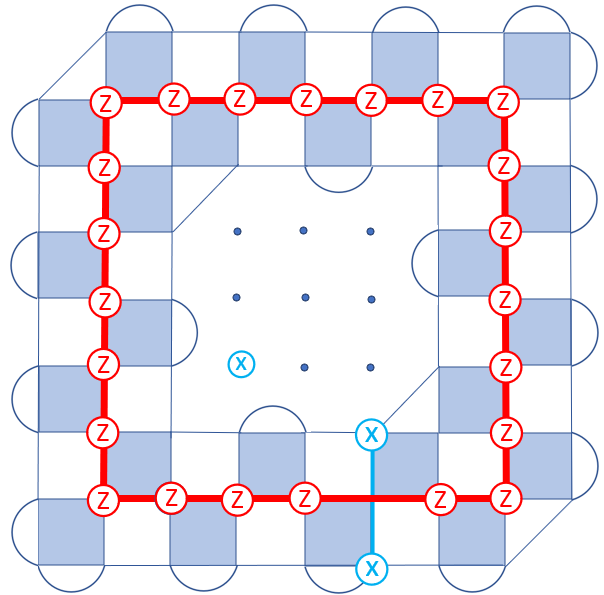}} 
    \subfigure(d){\includegraphics[width=0.192\textwidth]{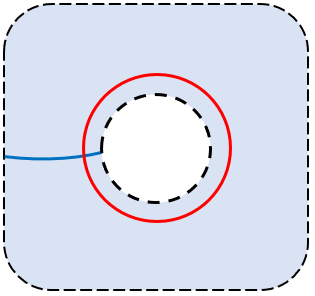}} 
    \caption{Different types of puncture defects on the toric code. The puncture and code boundary in (a) are rough while the puncture and code boundary in (c) are smooth. The measured stabilisers creating the punctures,  and non-contractible loops stabilising them are also shown for each puncture type in (a) and (c). Panels (b) and (c) show their respective diagrammatic representations as introduced in ref.~\cite{Brown_2017}.}
    \label{fig:puncture1}
\end{figure}

A useful way to encode information on the toric code with open boundary conditions, i.e. the planar code, is to introduce defects on its surface. One such defect is the puncture, which consists of a hole on the lattice created by measuring stabilisers so as to disentangle the measured spin systems from the code \cite{delfosse2016generalized, Fowler_2012}. The type of boundary of a puncture depends on which type of stabiliser was measured in its creation, namely rough (smooth) boundary for Pauli $Z (X)$-type as shown in Fig.~\ref{fig:puncture1}. When the code and puncture boundaries are of the same type, a logical qubit is encoded by defining a logical operator $\bar{X}$ as a sequence of Pauli-$X$ operations supported on qubits along a loop enclosing the puncture, and $\bar{Z}$ as a string of Pauli-$Z$ applied on qubits between code and puncture boundaries, satisfying the necessary anti-commutation as described in Fig.~\ref{fig:puncture1} (a) and (b), and equivalently for a smooth puncture in Fig.~\ref{fig:puncture1} (c) and (d). From the topological anyon picture, the two-level system is designed by encoding the parity of the puncture's anyon population, where each anyon has been passed from the code boundary to the puncture. These are $e$ anyons if the boundaries are rough, and $m$ if smooth. 

\begin{figure}
    \centering
    \subfigure(a){\includegraphics[width=0.19\textwidth]{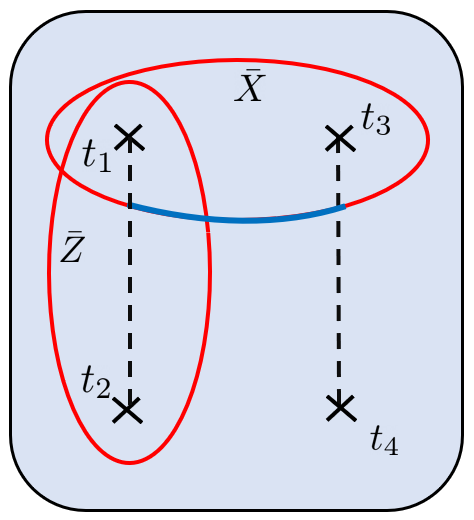}} 
    \subfigure(b){\includegraphics[width=0.19\textwidth]{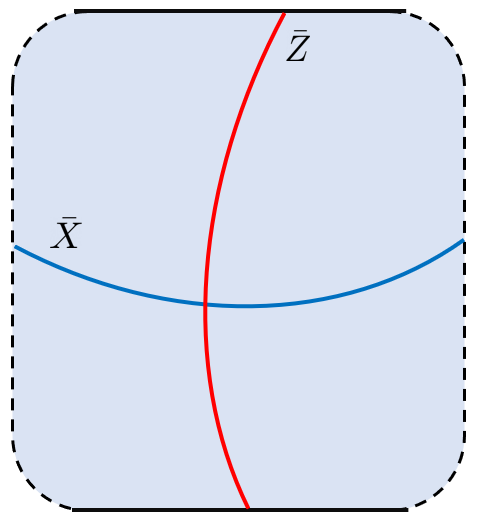}} 
    \caption{Twists on the toric code. Panel (a) shows  a qubit encoded using two pairs of twists, with logical operators $\bar{X}$ and $\bar{Z}$. Panel (b) shows twist defect lines moved to the corners of the code boundary \cite{Brown_2017}.}
    \label{fig:twists}
\end{figure}

Another type of extrinsic defects on the planar code are twists which are created by introducing a translation \cite{Bombin_2010} or a series of measurements on the lattice \cite{Brown_2017} modifying its stabilisers. As with MZMs, twists were shown to behave like Ising anyons. Two pairs of twists can encode a logical qubit as shown in Fig.~\ref{fig:twists}, and logical Pauli operations are achieved by braiding twists. In ref.~\cite{Brown_2017} Brown \textit{et} al. showed that the planar code with mixed boundaries supports corner defects which can be deformed into twists on the lattice. Hence, there is an equivalence between the right and left panels in Fig.~\ref{fig:twists}. Our graphical notation is consistent with the language introduced in ref.~\cite{Brown_2017}. The blue background represents the planar code bulk; a dashed line is a smooth boundary condensing $m$ anyons, while a continuous one is rough and condenses $e$ anyons. 

\section{Fusion and exchange of mixed-boundary punctures}
\label{sec:naStats}

\subsection{The system}

\begin{figure}
    \centering
    \includegraphics[width=0.20\textwidth]{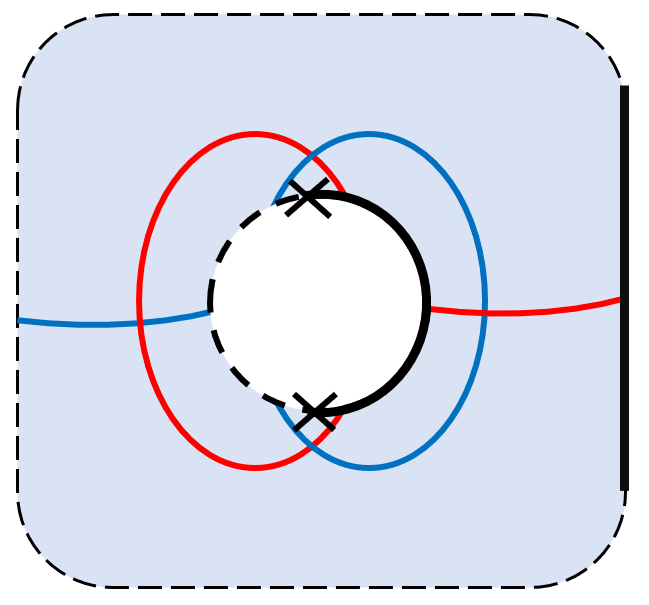}
    \caption{A mixed-boundary puncture. Both blue ($X$-type) and red ($Z$-type) strings can terminate at its boundaries. The red and blue loops stabilise this defect, and the crosses indicate the meeting point of rough and smooth boundaries, i.e. twists. Note that we need a hybrid code boundary for the attached strings.}
    \label{fig:mixed}
\end{figure}

In Fig.~\ref{fig:twists}(b) corners of the planar code correspond to points at which smooth and rough boundaries are juxtaposed. Given their relationship with twists, we ask whether punctures with mixed-boundaries can exhibit Ising-like behaviour. Indeed by this definition, one can see that a puncture with mixed boundaries, shown in Fig.~\ref{fig:mixed}, carries two twists, located at the meeting point of the different boundaries.

A mixed-boundary puncture is created by measuring both $X$ and $Z$-type stabilisers \cite{Bombin_2009, delfosse2016generalized}. The strings allowed at its boundaries, and the loop operators that stabilise it in Fig.~\ref{fig:mixed} indicate that a mixed-boundary puncture condenses both $e$ and $m$ anyons. Since one can encode a qubit using four MZMs or twists on the toric code, and achieve Clifford gates on its state through pair-wise braidings, our system will be composed of four copies of mixed boundary-punctures. However, despite the ability of these punctures to hold both toric code anyons, their braiding remains Abelian. Hence, we introduce non-locality in the encoding of Abelian anyons in order to generate the non-Abelian character. This is done by taking superpositions of anyons populating the punctures, which translates into superpositions of strings between each pair of punctures.

\subsection{Logical encoding}

\begin{figure}[t]
  \centering
  \advance\leftskip-7mm 
  \includegraphics[width=0.36\textwidth]{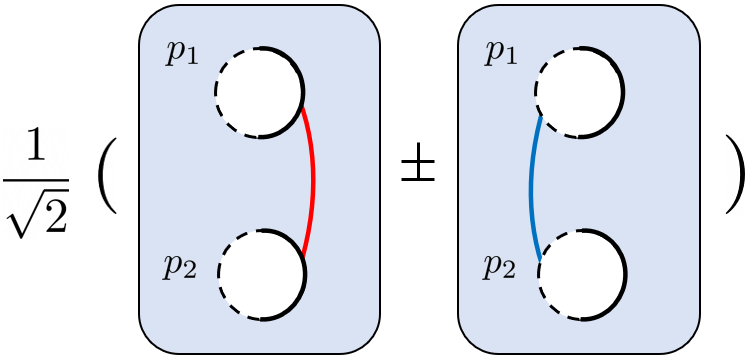}
  \caption{A state of a pair of mixed-boundary punctures. This state is defined in Eq.~\ref{eq:superposition} and describes a superposition of red and blue string configurations, respectivelym describing $p_1$ and $p_2$ each one absorbing an $e$ or an $m$ anyon.}
  \label{fig:superposition}
\end{figure}

We consider a pair of punctures with mixed boundaries created from vacuum, denoted by $p_1$ and $p_2$, and allow strings between their matching boundaries. We denote the state of a pair of punctures by its anyon population such that the state of a pair enclosing an $e$ anyon in each puncture is $\ket{ee}$; this corresponds to a red string with endpoints at each puncture, and likewise blue for $\ket{mm}$. Since the anyons are \textit{inside} the punctures, we remain in the ground state of the code as opposed to an open string which has excitations at its endpoints. We now let the pair ($p_1$,$p_2)$ be in the superposition of states given by
\begin{equation}
    \ket{p_1,p_2;\pm}=\frac{\ket{e_1e_2} \pm \ket{m_1m_2}}{\sqrt{2}}
    \label{eq:superposition}
\end{equation}
where this notation translates to the two-puncture system being in a superposition of red and blue string configurations, as shown in Fig.~\ref{fig:superposition}, and the states given by Eq.~\ref{eq:superposition} are degenerate. In fact, this choice of superposition is motivated by the fusion rules of Ising anyons in Eq.~\ref{eq:isingmodel}. 

\begin{figure}[t]
    \centering
    \subfigure(a){\includegraphics[width=0.195\textwidth]{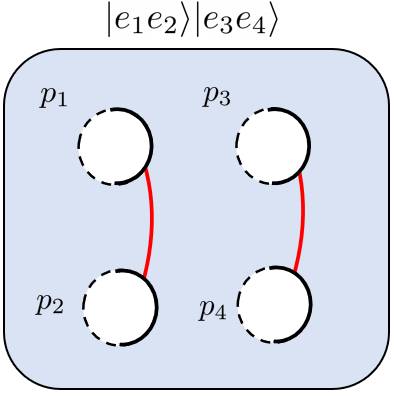}} 
    \subfigure(b){\includegraphics[width=0.195\textwidth]{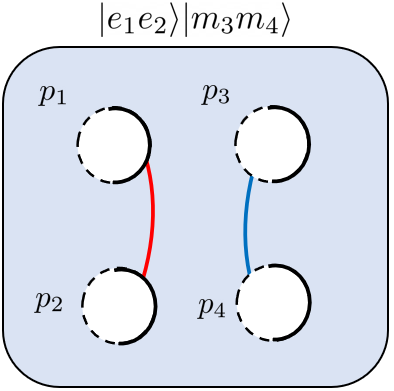}} 
    \subfigure(c){\includegraphics[width=0.195\textwidth]{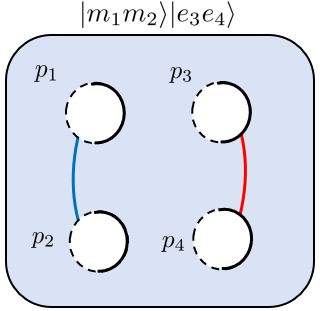}}
    \subfigure(d){\includegraphics[width=0.195\textwidth]{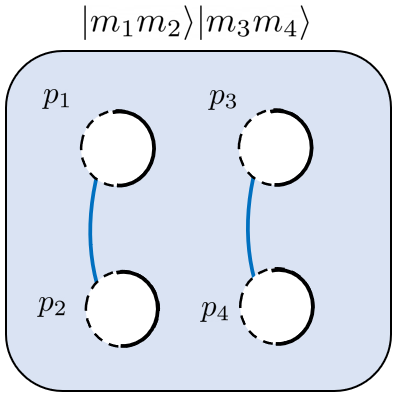}}
    \caption{Logical encoding of four mixed-boundary punctures. The system is separated into two pairs each in a state described in Fig.~\ref{fig:superposition}. The string configurations and their corresponding quantum states are based on puncture populations, and each quadrant represents a term in the joint state in Eq.~\ref{eq:jointState1}.}
    \label{fig:allStates}
\end{figure} 

We will consider two such pairs of punctures ($p_1$,$p_2$) and ($p_3,p_4$), and using Eq.~\ref{eq:superposition} write their general joint state as

\begin{equation}
\begin{aligned}
    \ket{( p_1,p_2;\pm )( p_3,p_4;\pm )} = {} & \\ \hspace{-10mm}\frac{1}{2}(\ket{e_1e_2}\ket{e_3e_4} &
    \pm  \ket{e_1e_2}\ket{m_3m_4} \\ & \hspace{-15mm}  \pm\ket{m_1m_2}\ket{e_3e_4}+\ket{m_1m_2}\ket{m_3m_4}).
    \label{eq:jointState1}
\end{aligned}
\end{equation}

Note that these configurations are constructed from local lattice defects where non-local quantum operations can be encoded. The terms in Eq.~\ref{eq:jointState1} correspond to the string configurations indicated in Fig.~\ref{fig:allStates}.

\subsection{Fusion}

We can verify that this system of punctures reproduces the fusion properties characteristic of Ising anyons by respectively fusing the charge contents of the pairs $(p_1,p_3)$ and $(p_2,p_4)$. Indeed, the fusion takes the joint state in Eq.~\ref{eq:jointState1} to the state
\begin{equation}
    \ket{p_{13},p_{24};\pm} = \frac{1}{\sqrt{2}}(\ket{1_{13},1_{24}} \pm \ket{\psi_{13},\psi_{24}})
    \label{eq:fusedState}
\end{equation}
where we define 
\begin{equation}
    \ket{1_{13},1_{24}} = \frac{1}{\sqrt{2}}(\ket{e_1e_2}\ket{e_3e_4}+\ket{m_1m_2}\ket{m_3m_4})
    \label{eq:vacuum}
\end{equation}
and
\begin{equation}
    \ket{\psi_{13},\psi_{24}} = \frac{1}{\sqrt{2}}(\ket{e_1e_2}\ket{m_3m_4}+\ket{m_1m_2}\ket{e_3e_4})
    \label{eq:psi}
\end{equation}
analogously to the scheme in ref.~\cite{horner}. We can understand this as the punctures from the terms in Eq.~\ref{eq:vacuum} behaving as the vacuum charge since each composite object is made up of either two $e$ or two $m$ anyons, while the terms in Eq.~\ref{eq:psi} each behave as a fermion string (i.e. both red and blue strings). If we identify the states $\ket{1_{13},1_{24}}$ and $\ket{\psi_{13},\psi_{24}}$ respectively with the vacuum and $\psi$ fermion sectors as described by the basis states in Eq.~\ref{fig:basissector}, then they are related to the fusion outcomes in Eq.~\ref{eq:fusedState} by a fusion matrix 

\begin{equation}
    F_{punct} = \frac{1}{\sqrt{2}}\begin{pmatrix}
        1 & 1 \\
        1 & -1
    \end{pmatrix}
    \label{eq:fusionfinale}
\end{equation}

which matches the Ising model fusion properties in Eq.~\ref{eq:fmatrix}. 


\subsection{Braiding}

We are now interested in how the state in Eq.~\ref{eq:jointState1} is affected by braiding individual punctures. For this purpose, we can encode a logical qubit using the configuration described in Eq.~\ref{eq:jointState1}, in the logical basis $\{\ket{++},\ket{--}\}$ where $\ket{++}=\ket{p_1,p_2;+}\ket{p_3,p_4;+}$ and $\ket{--}=\ket{p_1,p_2;-}\ket{p_3,p_4;-}$. This corresponds to the even parity sector. The basis in the odd parity sector is $\{\ket{+-},\ket{-+}\}$ but we will not consider it here. 

\begin{figure}
    \centering
    \subfigure(a){\includegraphics[width=0.205\textwidth]{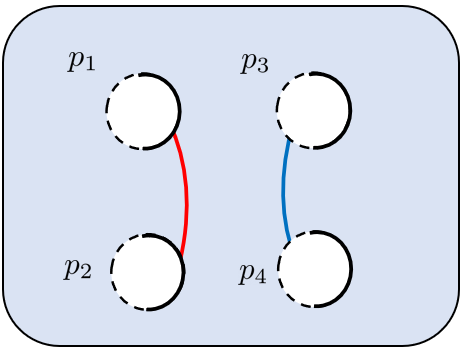}} 
    \subfigure(b){\includegraphics[width=0.204\textwidth]{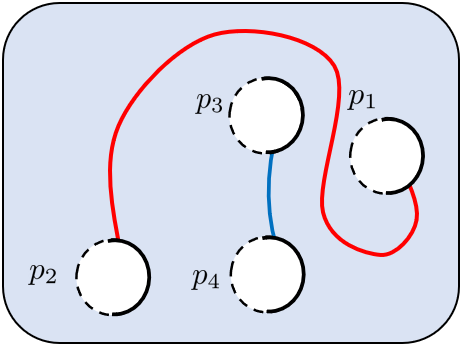}} 
    \subfigure(c){\includegraphics[width=0.205\textwidth]{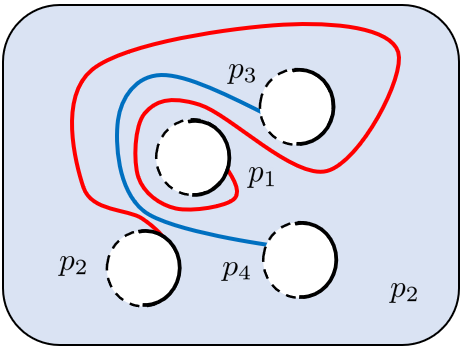}}
    \subfigure(d){\includegraphics[width=0.205\textwidth]{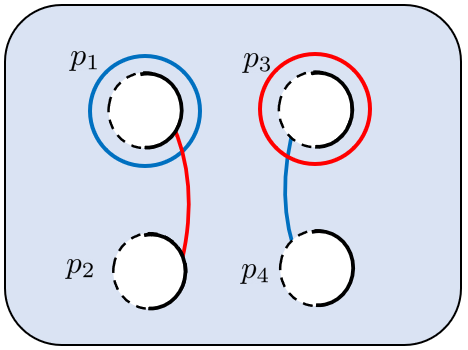}
    }
    \caption{Braiding operation shown for state $\ket{e_1e_2}\ket{m_3m_4}$. The steps are shown in (a) - (d), taking $p_1$ around $p_3$ for the string configuration in Fig.~\ref{fig:allStates}(c), and illustrate the full exchange of the $e$ anyon in $p_1$ around $m$ in $p_3$. The braiding has to be carried out without performing a self-twist of $p_1$ in order to recover (d).}
    \label{fig:foobar}
\end{figure}

Braiding $p_1$ around $p_3$ affects the states shown in Fig.~\ref{fig:allStates} differently. The case for $\ket{e_1e_2}\ket{m_3m_4}$ is detailed in Fig.~\ref{fig:foobar}, where the step between panels (c) and (d) consists of respectively multiplying the red and blue strings by a $Z$-type stabiliser (i.e. a red loop) and $X$-type stabiliser (i.e. a blue loop) operator, which are trivial operations on the toric code. We notice that in addition to the initial string configuration, punctures $p_1$ and $p_3$ are now enclosed by $X$ (blue) and $Z$ (red) loop 

\begin{figure}[t]
    \centering
    \subfigure(a){\includegraphics[width=0.195\textwidth]{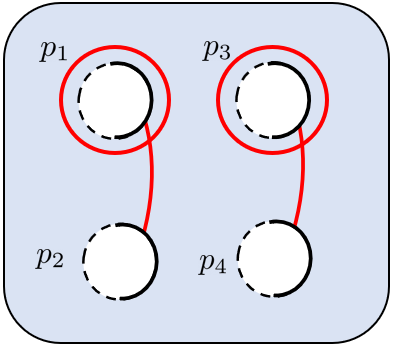}} 
    \subfigure(b){\includegraphics[width=0.195\textwidth]{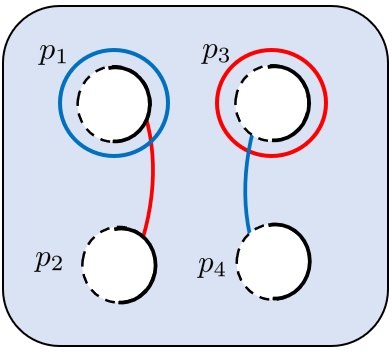}} 
    \subfigure(c){\includegraphics[width=0.195\textwidth]{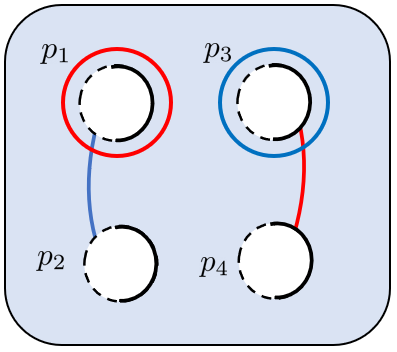}}
    \subfigure(d){\includegraphics[width=0.195\textwidth]{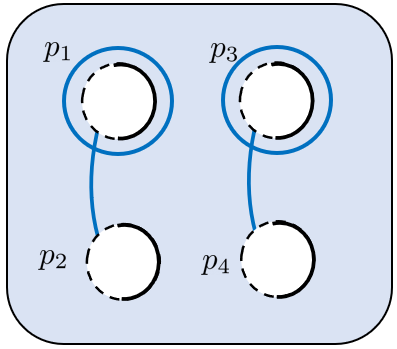}}
    \caption{Logical gate by braiding operation in the four-puncture system. In all panels $p_1$ and $p_3$ were exchanged. The evolution in (a) and (d) is trivial, but combined with (b) and (c) affects the non-local superposition state in  Eq.~\ref{eq:jointState1} (as shown in Fig.~\ref{fig:allStates}) non-trivially. This results in this final string configuration describing the state in  Eq.~\ref{eq:jointState2}.} 
    \label{fig:finalConfig}
\end{figure}

\hspace{-4mm}operators after the braiding, crossing the original strings which are of opposite-type. This evolution is a result of braiding the $e$ anyon in $p_1$ around the $m$ anyon in $p_3$. This is particularly interesting when considering how each term in Eq.~\ref{fig:allStates} evolves under the braiding. Indeed, we show the final configurations in Fig.~\ref{fig:finalConfig}, where only panels (b) and (c) have $X$ and $Z$ strings crossing and hence anti-commuting, while the braiding in (a) and (d) results in string crossings of the same type, i.e. Abelian. In fact the braiding in (a) and (b) panels is equivalent to full self-rotations of $p_1$ and $p_3$. The combined effect from this exchange acts on an encoded qubit non-trivially. Indeed, braiding $p_1$ around $p_3$ flips the sign in the second and third terms of Eq.~\ref{eq:jointState1} due to the mutual statistics of the toric code $e$ and $m$ anyons, resulting in the state in Eq.~\ref{eq:jointState2}:


\begin{equation}
\begin{aligned}
         \ket{(p_1,p_2;\mp)(p_3,p_4;\mp)} = {} & \\ \hspace{-10mm}\frac{1}{2}(\ket{e_1e_2}\ket{e_3e_4} & 
         \mp \ket{e_1e_2}\ket{m_3m_4} \\ &\hspace{-15mm} \mp \ket{m_1m_2}\ket{e_3e_4}+\ket{m_1m_2}\ket{m_3m_4}).
    \label{eq:jointState2}
\end{aligned}
\end{equation}

Upon rewriting Eq.~\ref{eq:jointState2} as a product of the states of pairs $(p_1,p_2)$ and $(p_3,p_4)$, one can see that the braiding changes the relative phase in the superpositions of both states. This transforms the logical encoding basis following
\begin{align}
B^2_{13}\ket{++} = \ket{--},
\\
B^2_{13}\ket{--} = \ket{++},
\label{eqn:eqlabel}
\end{align}

\begin{figure}[ht!]
    \centering
    \includegraphics[width=0.2\textwidth]{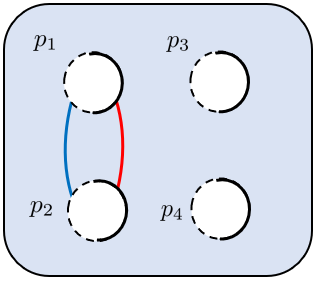}
    \caption{Logical $Z$ operator for the logical encoding described in Eq.~\ref{eq:jointState1} using the four mixed-boundary punctures.}
    \label{fig:Zgate}
\end{figure}

\hspace{-4mm}where $B_{23}^2$ denotes the full braid of $p_1$ around $p_3$. This is identified with the Pauli-$X$ operation, which is the signature of non-Abelian statistics of Ising anyon exchange given in Eq.~\ref{eq:anyonX}. 
We also observe from Eq.~\ref{eq:jointState1} that braiding $p_1$ and $p_4$ or alternatively $p_2$ and $p_3$ or $p_2$ and $p_4$ changes the state in an equivalent fashion. However, braiding $p_1$ and $p_2$ or $p_3$ and $p_4$ (ie. punctures from the same pair) acts trivially on Eq.~\ref{eq:jointState1} and likewise on our logical basis. Therefore it appears that we cannot recreate the full set of operations achievable with Ising anyons (and twists by association). Indeed, obtaining a Pauli-$Z$ operation in the same basis requires an operation that transform states according to: $\ket{e_1e_2} \xrightarrow{} \ket{m_1m_2} $ and $\ket{m_1m_2} \xrightarrow{} \ket{e_1e_2}$, which cannot be done exclusively by braiding operations in our system. We identify this $Z$ logical operator with the string combination in Fig.~\ref{fig:Zgate}. In consequence, the above encoding does not benefit from the simplicity of the logical operators available with twists since the logical $X$, which corresponds to applying the loop superposition in Fig.~\ref{fig:finalConfig}, and $Z$ and cannot be interchanged by puncture braiding only.

\hspace{2mm}

\section{Discussion}
\label{sec:conclusion}

We studied an unusual defect on the toric code referred to as mixed-boundary puncture, which is introduced by two types of stabiliser measurements creating a hybrid boundary. This opened interesting possibilities of anyon configurations combined with the punctures. It was recently shown that twists in the non-Abelian phase of the KHLM not only localise Majorana zero-modes but are also equivalent to vortices via an adiabatic lattice transformation \cite{horner}. Motivated by this we explored such an equivalence between twists and punctures using code deformations, in an attempt to find a local defect on the toric code that localises MZMs. However, we found no such adiabatic transformation. Indeed, in the Abelian phase of the KHM twists and punctures behave differently in that only twists have the ability to act as sources and sinks for $\psi$ fermions and recreate the Ising anyon fusion space. In contrast, while punctures ensure fault-tolerance, regardless of their boundary composition they cannot provide the mechanical non-locality offered by twists. In fact, mixed-boundary punctures inherently support a pair of twists on their boundary, and we find that any adiabatic mechanism that would promote them to an object that behaves as an Ising anyon requires moving the twists from the puncture boundary to the bulk of the code, which undermines the integrity of the puncture. Therefore, in order to recover the non-Abelian statistics on the toric code, we introduced a scheme that embraces the non-locality of twists by simulating their domain wall using our logical encoding. Indeed, we considered a logical basis for computation formulated on a superposition of the anyonic population of four mixed-boundary punctures, and found that braiding punctures from distinct pairs created from vacuum induces a Pauli-$X$ operation on the encoded qubit, thus reproducing the non-Abelian exchange statistics characteristic of the Ising model. This type of structure appears to be the way around using twist defects in order to recreate Majorana statistics on the toric code. 

The existing schemes such as those presented in Refs.~\cite{Bombin_2009, delfosse2016generalized} also utilise punctures to realise fault-tolerant gates by braiding, using combinations of $X$ and $Z$ boundary configurations, and can achieve Clifford and entangling gates. In contrast, further braiding operations with our chosen encoding do not expand our gate-set to the full scope of logical operations accessible with twist and MZM exchange, and therefore does not recover the full set of Clifford gates topologically. However, we emphasize that the aforementioned operations are
produced by the Abelian braiding of toric code anyons, while we exploit similar defects, with a different encoding combined with braiding to give rise to non-Abelian statistics.



\begin{acknowledgments}
We would like to thank Tom R. Scruby and Benjamin J. Brown for insightful discussions and explanations. A.B. acknowledges funding from the EPSRC Centre for Doctoral Training in Delivering Quantum Technologies at UCL, Grant No. EP/S021582/1L, D.E.B. was supported by the Quantera project Quantum Codes Design and Architecture EPSRC grant EP/R043647/1, and J.K.P. by EPSRC Grant No. EP/R020612/1.
\end{acknowledgments}

\nocite{*}

\bibliography{mybibliography.bib}

\begin{thebibliography}{27}%
\makeatletter
\providecommand \@ifxundefined [1]{%
 \@ifx{#1\undefined}
}%
\providecommand \@ifnum [1]{%
 \ifnum #1\expandafter \@firstoftwo
 \else \expandafter \@secondoftwo
 \fi
}%
\providecommand \@ifx [1]{%
 \ifx #1\expandafter \@firstoftwo
 \else \expandafter \@secondoftwo
 \fi
}%
\providecommand \natexlab [1]{#1}%
\providecommand \enquote  [1]{``#1''}%
\providecommand \bibnamefont  [1]{#1}%
\providecommand \bibfnamefont [1]{#1}%
\providecommand \citenamefont [1]{#1}%
\providecommand \href@noop [0]{\@secondoftwo}%
\providecommand \href [0]{\begingroup \@sanitize@url \@href}%
\providecommand \@href[1]{\@@startlink{#1}\@@href}%
\providecommand \@@href[1]{\endgroup#1\@@endlink}%
\providecommand \@sanitize@url [0]{\catcode `\\12\catcode `\$12\catcode
  `\&12\catcode `\#12\catcode `\^12\catcode `\_12\catcode `\%12\relax}%
\providecommand \@@startlink[1]{}%
\providecommand \@@endlink[0]{}%
\providecommand \url  [0]{\begingroup\@sanitize@url \@url }%
\providecommand \@url [1]{\endgroup\@href {#1}{\urlprefix }}%
\providecommand \urlprefix  [0]{URL }%
\providecommand \Eprint [0]{\href }%
\providecommand \doibase [0]{https://doi.org/}%
\providecommand \selectlanguage [0]{\@gobble}%
\providecommand \bibinfo  [0]{\@secondoftwo}%
\providecommand \bibfield  [0]{\@secondoftwo}%
\providecommand \translation [1]{[#1]}%
\providecommand \BibitemOpen [0]{}%
\providecommand \bibitemStop [0]{}%
\providecommand \bibitemNoStop [0]{.\EOS\space}%
\providecommand \EOS [0]{\spacefactor3000\relax}%
\providecommand \BibitemShut  [1]{\csname bibitem#1\endcsname}%
\let\auto@bib@innerbib\@empty
\bibitem [{\citenamefont {Bombin}(2010)}]{Bombin_2010}%
  \BibitemOpen
  \bibfield  {author} {\bibinfo {author} {\bibfnamefont {H.}~\bibnamefont
  {Bombin}},\ }\bibfield  {title} {\bibinfo {title} {Topological order with a
  twist: Ising anyons from an {A}belian model},\ }\bibfield  {journal}
  {\bibinfo  {journal} {Physical Review Letters}\ }\textbf {\bibinfo {volume}
  {105}},\ \href {https://doi.org/10.1103/physrevlett.105.030403}
  {10.1103/physrevlett.105.030403} (\bibinfo {year} {2010})\BibitemShut
  {NoStop}%
\bibitem [{\citenamefont {{W}ilczek}(1982)}]{PhysRevLett.49.957}%
  \BibitemOpen
  \bibfield  {author} {\bibinfo {author} {\bibfnamefont {F.}~\bibnamefont
  {{W}ilczek}},\ }\bibfield  {title} {\bibinfo {title} {Quantum mechanics of
  fractional-spin particles},\ }\href
  {https://doi.org/10.1103/PhysRevLett.49.957} {\bibfield  {journal} {\bibinfo
  {journal} {Phys. Rev. Lett.}\ }\textbf {\bibinfo {volume} {49}},\ \bibinfo
  {pages} {957} (\bibinfo {year} {1982})}\BibitemShut {NoStop}%
\bibitem [{\citenamefont {{F}reedman}\ \emph {et~al.}(2002)\citenamefont
  {{F}reedman}, \citenamefont {{L}arsen},\ and\ \citenamefont
  {{W}ang}}]{Freedman}%
  \BibitemOpen
  \bibfield  {author} {\bibinfo {author} {\bibfnamefont {M.}~\bibnamefont
  {{F}reedman}}, \bibinfo {author} {\bibfnamefont {M.}~\bibnamefont
  {{L}arsen}},\ and\ \bibinfo {author} {\bibfnamefont {Z.}~\bibnamefont
  {{W}ang}},\ }\bibfield  {title} {\bibinfo {title} {A modular functor which is
  universal for quantum computation},\ }\href@noop {} {\bibfield  {journal}
  {\bibinfo  {journal} {Commun. Math. Phys.}\ ,\ \bibinfo {pages} {605–622}}
  (\bibinfo {year} {2002})}\BibitemShut {NoStop}%
\bibitem [{\citenamefont {Pachos}(2012)}]{pachos_2012}%
  \BibitemOpen
  \bibfield  {author} {\bibinfo {author} {\bibfnamefont {J.~K.}\ \bibnamefont
  {Pachos}},\ }\href {https://doi.org/10.1017/CBO9780511792908} {\emph
  {\bibinfo {title} {Introduction to Topological Quantum Computation}}}\
  (\bibinfo  {publisher} {Cambridge University Press},\ \bibinfo {year}
  {2012})\BibitemShut {NoStop}%
\bibitem [{\citenamefont {{O}gburn}\ \emph {et~al.}(1999)\citenamefont
  {{O}gburn}, ,\ and\ \citenamefont {{P}reskill}}]{preskill}%
  \BibitemOpen
  \bibfield  {author} {\bibinfo {author} {\bibfnamefont {R.~W.}\ \bibnamefont
  {{O}gburn}}, ,\ and\ \bibinfo {author} {\bibfnamefont {J.}~\bibnamefont
  {{P}reskill}},\ }\bibfield  {title} {\bibinfo {title} {Topological quantum
  computation},\ }in\ \href@noop {} {\emph {\bibinfo {booktitle} {Quantum
  Computing and Quantum Communications}}},\ \bibinfo {editor} {edited by\
  \bibinfo {editor} {\bibfnamefont {C.~P.}\ \bibnamefont {{W}illiams}}}\
  (\bibinfo  {publisher} {Springer Berlin Heidelberg},\ \bibinfo {address}
  {Berlin, Heidelberg},\ \bibinfo {year} {1999})\ pp.\ \bibinfo {pages}
  {341--356}\BibitemShut {NoStop}%
\bibitem [{\citenamefont {{K}itaev}(2003)}]{Kitaev_2003}%
  \BibitemOpen
  \bibfield  {author} {\bibinfo {author} {\bibfnamefont {A.}~\bibnamefont
  {{K}itaev}},\ }\bibfield  {title} {\bibinfo {title} {Fault-tolerant quantum
  computation by anyons},\ }\href
  {https://doi.org/10.1016/s0003-4916(02)00018-0} {\bibfield  {journal}
  {\bibinfo  {journal} {Annals of Physics}\ }\textbf {\bibinfo {volume}
  {303}},\ \bibinfo {pages} {2–30} (\bibinfo {year} {2003})}\BibitemShut
  {NoStop}%
\bibitem [{\citenamefont {Kitaev}(2001)}]{Kitaev_2001}%
  \BibitemOpen
  \bibfield  {author} {\bibinfo {author} {\bibfnamefont {A.~Y.}\ \bibnamefont
  {Kitaev}},\ }\bibfield  {title} {\bibinfo {title} {Unpaired majorana fermions
  in quantum wires},\ }\href {https://doi.org/10.1070/1063-7869/44/10s/s29}
  {\bibfield  {journal} {\bibinfo  {journal} {Physics-Uspekhi}\ }\textbf
  {\bibinfo {volume} {44}},\ \bibinfo {pages} {131–136} (\bibinfo {year}
  {2001})}\BibitemShut {NoStop}%
\bibitem [{\citenamefont {Fu}\ and\ \citenamefont {Kane}(2008)}]{Fu_2008}%
  \BibitemOpen
  \bibfield  {author} {\bibinfo {author} {\bibfnamefont {L.}~\bibnamefont
  {Fu}}\ and\ \bibinfo {author} {\bibfnamefont {C.~L.}\ \bibnamefont {Kane}},\
  }\bibfield  {title} {\bibinfo {title} {Superconducting proximity effect and
  majorana fermions at the surface of a topological insulator},\ }\bibfield
  {journal} {\bibinfo  {journal} {Physical Review Letters}\ }\textbf {\bibinfo
  {volume} {100}},\ \href {https://doi.org/10.1103/physrevlett.100.096407}
  {10.1103/physrevlett.100.096407} (\bibinfo {year} {2008})\BibitemShut
  {NoStop}%
\bibitem [{\citenamefont {Gottesman}(1997)}]{phdthesis}%
  \BibitemOpen
  \bibfield  {author} {\bibinfo {author} {\bibfnamefont {D.}~\bibnamefont
  {Gottesman}},\ }\emph {\bibinfo {title} {Stabilizer Codes and Quantum Error
  Correction}},\ \href@noop {} {Ph.D. thesis},\ \bibinfo  {school} {California
  Institute of Technology} (\bibinfo {year} {1997})\BibitemShut {NoStop}%
\bibitem [{\citenamefont {Fowler}\ \emph {et~al.}(2012)\citenamefont {Fowler},
  \citenamefont {Mariantoni}, \citenamefont {Martinis},\ and\ \citenamefont
  {Cleland}}]{Fowler_2012}%
  \BibitemOpen
  \bibfield  {author} {\bibinfo {author} {\bibfnamefont {A.~G.}\ \bibnamefont
  {Fowler}}, \bibinfo {author} {\bibfnamefont {M.}~\bibnamefont {Mariantoni}},
  \bibinfo {author} {\bibfnamefont {J.~M.}\ \bibnamefont {Martinis}},\ and\
  \bibinfo {author} {\bibfnamefont {A.~N.}\ \bibnamefont {Cleland}},\
  }\bibfield  {title} {\bibinfo {title} {Surface codes: Towards practical
  large-scale quantum computation},\ }\bibfield  {journal} {\bibinfo  {journal}
  {Physical Review A}\ }\textbf {\bibinfo {volume} {86}},\ \href
  {https://doi.org/10.1103/physreva.86.032324} {10.1103/physreva.86.032324}
  (\bibinfo {year} {2012})\BibitemShut {NoStop}%
\bibitem [{\citenamefont {Kitaev}(2006)}]{Kitaev_2006}%
  \BibitemOpen
  \bibfield  {author} {\bibinfo {author} {\bibfnamefont {A.}~\bibnamefont
  {Kitaev}},\ }\bibfield  {title} {\bibinfo {title} {Anyons in an exactly
  solved model and beyond},\ }\href {https://doi.org/10.1016/j.aop.2005.10.005}
  {\bibfield  {journal} {\bibinfo  {journal} {Annals of Physics}\ }\textbf
  {\bibinfo {volume} {321}},\ \bibinfo {pages} {2–111} (\bibinfo {year}
  {2006})}\BibitemShut {NoStop}%
\bibitem [{\citenamefont {Kells}\ \emph {et~al.}(2009)\citenamefont {Kells},
  \citenamefont {Slingerland},\ and\ \citenamefont {Vala}}]{Kells_2009}%
  \BibitemOpen
  \bibfield  {author} {\bibinfo {author} {\bibfnamefont {G.}~\bibnamefont
  {Kells}}, \bibinfo {author} {\bibfnamefont {J.~K.}\ \bibnamefont
  {Slingerland}},\ and\ \bibinfo {author} {\bibfnamefont {J.}~\bibnamefont
  {Vala}},\ }\bibfield  {title} {\bibinfo {title} {Description of kitaev’s
  honeycomb model with toric-code stabilizers},\ }\bibfield  {journal}
  {\bibinfo  {journal} {Physical Review B}\ }\textbf {\bibinfo {volume} {80}},\
  \href {https://doi.org/10.1103/physrevb.80.125415}
  {10.1103/physrevb.80.125415} (\bibinfo {year} {2009})\BibitemShut {NoStop}%
\bibitem [{\citenamefont {Dennis}\ \emph {et~al.}(2002)\citenamefont {Dennis},
  \citenamefont {Kitaev}, \citenamefont {Landahl},\ and\ \citenamefont
  {Preskill}}]{Dennis_2002}%
  \BibitemOpen
  \bibfield  {author} {\bibinfo {author} {\bibfnamefont {E.}~\bibnamefont
  {Dennis}}, \bibinfo {author} {\bibfnamefont {A.}~\bibnamefont {Kitaev}},
  \bibinfo {author} {\bibfnamefont {A.}~\bibnamefont {Landahl}},\ and\ \bibinfo
  {author} {\bibfnamefont {J.}~\bibnamefont {Preskill}},\ }\bibfield  {title}
  {\bibinfo {title} {Topological quantum memory},\ }\href
  {https://doi.org/10.1063/1.1499754} {\bibfield  {journal} {\bibinfo
  {journal} {Journal of Mathematical Physics}\ }\textbf {\bibinfo {volume}
  {43}},\ \bibinfo {pages} {4452–4505} (\bibinfo {year} {2002})}\BibitemShut
  {NoStop}%
\bibitem [{\citenamefont {Delfosse}\ \emph {et~al.}(2016)\citenamefont
  {Delfosse}, \citenamefont {Iyer},\ and\ \citenamefont
  {Poulin}}]{delfosse2016generalized}%
  \BibitemOpen
  \bibfield  {author} {\bibinfo {author} {\bibfnamefont {N.}~\bibnamefont
  {Delfosse}}, \bibinfo {author} {\bibfnamefont {P.}~\bibnamefont {Iyer}},\
  and\ \bibinfo {author} {\bibfnamefont {D.}~\bibnamefont {Poulin}},\
  }\href@noop {} {\bibinfo {title} {Generalized surface codes and packing of
  logical qubits}} (\bibinfo {year} {2016}),\ \Eprint
  {https://arxiv.org/abs/1606.07116} {arXiv:1606.07116 [quant-ph]} \BibitemShut
  {NoStop}%
\bibitem [{\citenamefont {Raussendorf}\ \emph {et~al.}(2007)\citenamefont
  {Raussendorf}, \citenamefont {Harrington},\ and\ \citenamefont
  {Goyal}}]{Raussendorf_2007}%
  \BibitemOpen
  \bibfield  {author} {\bibinfo {author} {\bibfnamefont {R.}~\bibnamefont
  {Raussendorf}}, \bibinfo {author} {\bibfnamefont {J.}~\bibnamefont
  {Harrington}},\ and\ \bibinfo {author} {\bibfnamefont {K.}~\bibnamefont
  {Goyal}},\ }\bibfield  {title} {\bibinfo {title} {Topological fault-tolerance
  in cluster state quantum computation},\ }\href
  {https://doi.org/10.1088/1367-2630/9/6/199} {\bibfield  {journal} {\bibinfo
  {journal} {New Journal of Physics}\ }\textbf {\bibinfo {volume} {9}},\
  \bibinfo {pages} {199–199} (\bibinfo {year} {2007})}\BibitemShut {NoStop}%
\bibitem [{\citenamefont {Barkeshli}\ \emph {et~al.}(2013)\citenamefont
  {Barkeshli}, \citenamefont {Jian},\ and\ \citenamefont
  {Qi}}]{Barkeshli_2013}%
  \BibitemOpen
  \bibfield  {author} {\bibinfo {author} {\bibfnamefont {M.}~\bibnamefont
  {Barkeshli}}, \bibinfo {author} {\bibfnamefont {C.-M.}\ \bibnamefont
  {Jian}},\ and\ \bibinfo {author} {\bibfnamefont {X.-L.}\ \bibnamefont {Qi}},\
  }\bibfield  {title} {\bibinfo {title} {Twist defects and projective
  non-{A}belian braiding statistics},\ }\bibfield  {journal} {\bibinfo
  {journal} {Physical Review B}\ }\textbf {\bibinfo {volume} {87}},\ \href
  {https://doi.org/10.1103/physrevb.87.045130} {10.1103/physrevb.87.045130}
  (\bibinfo {year} {2013})\BibitemShut {NoStop}%
\bibitem [{\citenamefont {Teo}(2016)}]{Teo_2016}%
  \BibitemOpen
  \bibfield  {author} {\bibinfo {author} {\bibfnamefont {J.~C.~Y.}\
  \bibnamefont {Teo}},\ }\bibfield  {title} {\bibinfo {title} {Globally
  symmetric topological phase: from anyonic symmetry to twist defect},\ }\href
  {https://doi.org/10.1088/0953-8984/28/14/143001} {\bibfield  {journal}
  {\bibinfo  {journal} {Journal of Physics: Condensed Matter}\ }\textbf
  {\bibinfo {volume} {28}},\ \bibinfo {pages} {143001} (\bibinfo {year}
  {2016})}\BibitemShut {NoStop}%
\bibitem [{\citenamefont {Zheng}\ \emph {et~al.}(2015)\citenamefont {Zheng},
  \citenamefont {Dua},\ and\ \citenamefont {Jiang}}]{Zheng_2015}%
  \BibitemOpen
  \bibfield  {author} {\bibinfo {author} {\bibfnamefont {H.}~\bibnamefont
  {Zheng}}, \bibinfo {author} {\bibfnamefont {A.}~\bibnamefont {Dua}},\ and\
  \bibinfo {author} {\bibfnamefont {L.}~\bibnamefont {Jiang}},\ }\bibfield
  {title} {\bibinfo {title} {Demonstrating non-abelian statistics of majorana
  fermions using twist defects},\ }\bibfield  {journal} {\bibinfo  {journal}
  {Physical Review B}\ }\textbf {\bibinfo {volume} {92}},\ \href
  {https://doi.org/10.1103/physrevb.92.245139} {10.1103/physrevb.92.245139}
  (\bibinfo {year} {2015})\BibitemShut {NoStop}%
\bibitem [{\citenamefont {You}\ and\ \citenamefont {Wen}(2012)}]{You_2012}%
  \BibitemOpen
  \bibfield  {author} {\bibinfo {author} {\bibfnamefont {Y.-Z.}\ \bibnamefont
  {You}}\ and\ \bibinfo {author} {\bibfnamefont {X.-G.}\ \bibnamefont {Wen}},\
  }\bibfield  {title} {\bibinfo {title} {zhuang},\ }\bibfield  {journal}
  {\bibinfo  {journal} {Physical Review B}\ }\textbf {\bibinfo {volume} {86}},\
  \href {https://doi.org/10.1103/physrevb.86.161107}
  {10.1103/physrevb.86.161107} (\bibinfo {year} {2012})\BibitemShut {NoStop}%
\bibitem [{\citenamefont {Krishna}\ and\ \citenamefont
  {Poulin}(2020)}]{Krishna_2020}%
  \BibitemOpen
  \bibfield  {author} {\bibinfo {author} {\bibfnamefont {A.}~\bibnamefont
  {Krishna}}\ and\ \bibinfo {author} {\bibfnamefont {D.}~\bibnamefont
  {Poulin}},\ }\bibfield  {title} {\bibinfo {title} {Topological wormholes:
  Nonlocal defects on the toric code},\ }\bibfield  {journal} {\bibinfo
  {journal} {Physical Review Research}\ }\textbf {\bibinfo {volume} {2}},\
  \href {https://doi.org/10.1103/physrevresearch.2.023116}
  {10.1103/physrevresearch.2.023116} (\bibinfo {year} {2020})\BibitemShut
  {NoStop}%
\bibitem [{\citenamefont {Horner}\ \emph {et~al.}(2020)\citenamefont {Horner},
  \citenamefont {Farjami},\ and\ \citenamefont {Pachos}}]{horner}%
  \BibitemOpen
  \bibfield  {author} {\bibinfo {author} {\bibfnamefont {M.~D.}\ \bibnamefont
  {Horner}}, \bibinfo {author} {\bibfnamefont {A.}~\bibnamefont {Farjami}},\
  and\ \bibinfo {author} {\bibfnamefont {J.~K.}\ \bibnamefont {Pachos}},\
  }\bibfield  {title} {\bibinfo {title} {Equivalence between vortices, twists,
  and chiral gauge fields in the {K}itaev honeycomb lattice model},\ }\href
  {https://doi.org/10.1103/PhysRevB.102.125152} {\bibfield  {journal} {\bibinfo
   {journal} {Phys. Rev. B}\ }\textbf {\bibinfo {volume} {102}},\ \bibinfo
  {pages} {125152} (\bibinfo {year} {2020})}\BibitemShut {NoStop}%
\bibitem [{\citenamefont {Wootton}\ \emph {et~al.}(2008)\citenamefont
  {Wootton}, \citenamefont {Lahtinen}, \citenamefont {Wang},\ and\
  \citenamefont {Pachos}}]{Wootton_2008}%
  \BibitemOpen
  \bibfield  {author} {\bibinfo {author} {\bibfnamefont {J.~R.}\ \bibnamefont
  {Wootton}}, \bibinfo {author} {\bibfnamefont {V.}~\bibnamefont {Lahtinen}},
  \bibinfo {author} {\bibfnamefont {Z.}~\bibnamefont {Wang}},\ and\ \bibinfo
  {author} {\bibfnamefont {J.~K.}\ \bibnamefont {Pachos}},\ }\bibfield  {title}
  {\bibinfo {title} {Non-{A}belian statistics from an {A}belian model},\
  }\bibfield  {journal} {\bibinfo  {journal} {Physical Review B}\ }\textbf
  {\bibinfo {volume} {78}},\ \href {https://doi.org/10.1103/physrevb.78.161102}
  {10.1103/physrevb.78.161102} (\bibinfo {year} {2008})\BibitemShut {NoStop}%
\bibitem [{\citenamefont {Ivanov}(2001)}]{ivanov}%
  \BibitemOpen
  \bibfield  {author} {\bibinfo {author} {\bibfnamefont {D.~A.}\ \bibnamefont
  {Ivanov}},\ }\bibfield  {title} {\bibinfo {title} {Non-{A}belian statistics
  of half-quantum vortices in $\mathit{p}$-wave superconductors},\ }\href
  {https://doi.org/10.1103/PhysRevLett.86.268} {\bibfield  {journal} {\bibinfo
  {journal} {Phys. Rev. Lett.}\ }\textbf {\bibinfo {volume} {86}},\ \bibinfo
  {pages} {268} (\bibinfo {year} {2001})}\BibitemShut {NoStop}%
\bibitem [{\citenamefont {Litinski}\ and\ \citenamefont {von
  Oppen}(2018)}]{Litinski_2018}%
  \BibitemOpen
  \bibfield  {author} {\bibinfo {author} {\bibfnamefont {D.}~\bibnamefont
  {Litinski}}\ and\ \bibinfo {author} {\bibfnamefont {F.}~\bibnamefont {von
  Oppen}},\ }\bibfield  {title} {\bibinfo {title} {Quantum computing with
  majorana fermion codes},\ }\bibfield  {journal} {\bibinfo  {journal}
  {Physical Review B}\ }\textbf {\bibinfo {volume} {97}},\ \href
  {https://doi.org/10.1103/physrevb.97.205404} {10.1103/physrevb.97.205404}
  (\bibinfo {year} {2018})\BibitemShut {NoStop}%
\bibitem [{\citenamefont {Brown}\ \emph {et~al.}(2017)\citenamefont {Brown},
  \citenamefont {Laubscher}, \citenamefont {Kesselring},\ and\ \citenamefont
  {Wootton}}]{Brown_2017}%
  \BibitemOpen
  \bibfield  {author} {\bibinfo {author} {\bibfnamefont {B.~J.}\ \bibnamefont
  {Brown}}, \bibinfo {author} {\bibfnamefont {K.}~\bibnamefont {Laubscher}},
  \bibinfo {author} {\bibfnamefont {M.~S.}\ \bibnamefont {Kesselring}},\ and\
  \bibinfo {author} {\bibfnamefont {J.~R.}\ \bibnamefont {Wootton}},\
  }\bibfield  {title} {\bibinfo {title} {Poking holes and cutting corners to
  achieve {C}lifford gates with the surface code},\ }\bibfield  {journal}
  {\bibinfo  {journal} {Physical Review X}\ }\textbf {\bibinfo {volume} {7}},\
  \href {https://doi.org/10.1103/physrevx.7.021029} {10.1103/physrevx.7.021029}
  (\bibinfo {year} {2017})\BibitemShut {NoStop}%
\bibitem [{\citenamefont {Bombin}\ and\ \citenamefont
  {Martin-Delgado}(2009)}]{Bombin_2009}%
  \BibitemOpen
  \bibfield  {author} {\bibinfo {author} {\bibfnamefont {H.}~\bibnamefont
  {Bombin}}\ and\ \bibinfo {author} {\bibfnamefont {M.~A.}\ \bibnamefont
  {Martin-Delgado}},\ }\bibfield  {title} {\bibinfo {title} {Quantum
  measurements and gates by code deformation},\ }\href
  {https://doi.org/10.1088/1751-8113/42/9/095302} {\bibfield  {journal}
  {\bibinfo  {journal} {Journal of Physics A: Mathematical and Theoretical}\
  }\textbf {\bibinfo {volume} {42}},\ \bibinfo {pages} {095302} (\bibinfo
  {year} {2009})}\BibitemShut {NoStop}%
\bibitem [{\citenamefont {Łodyga}\ \emph {et~al.}(2015)\citenamefont
  {Łodyga}, \citenamefont {Mazurek}, \citenamefont {Grudka},\ and\
  \citenamefont {Horodecki}}]{_odyga_2015}%
  \BibitemOpen
  \bibfield  {author} {\bibinfo {author} {\bibfnamefont {J.}~\bibnamefont
  {Łodyga}}, \bibinfo {author} {\bibfnamefont {P.}~\bibnamefont {Mazurek}},
  \bibinfo {author} {\bibfnamefont {A.}~\bibnamefont {Grudka}},\ and\ \bibinfo
  {author} {\bibfnamefont {M.}~\bibnamefont {Horodecki}},\ }\bibfield  {title}
  {\bibinfo {title} {Simple scheme for encoding and decoding a qubit in unknown
  state for various topological codes},\ }\bibfield  {journal} {\bibinfo
  {journal} {Scientific Reports}\ }\textbf {\bibinfo {volume} {5}},\ \href
  {https://doi.org/10.1038/srep08975} {10.1038/srep08975} (\bibinfo {year}
  {2015})\BibitemShut {NoStop}%
\end{thebibliography}%
\end{document}